\documentclass[11pt]{article}

\usepackage{amsmath,amssymb}
\usepackage{graphicx}
\usepackage{bm}
\usepackage{amsfonts}
\usepackage{algorithm, algorithmic}
\usepackage{subcaption}
\usepackage{here}
\usepackage{url}
\usepackage{fullpage}
\usepackage{authblk}
\usepackage{tabularx}
\usepackage{multirow}
\usepackage{booktabs}
\usepackage{tikz}
\usetikzlibrary{decorations.pathmorphing}
\usetikzlibrary{arrows}
\usetikzlibrary{positioning}
\usepackage{fancyhdr}

\fancypagestyle{FirstPage}{
\lfoot{{\footnotesize This document is the unedited Author's version of a Submitted Work that was subsequently accepted for publication in The Journal of Physical Chemistry A, copyright \copyright\ 2022 American Chemical Society after peer review. To access the final edited and published work see \url{https://pubs.acs.org/articlesonrequest/AOR-Q8FIMKQPDUTCHMFFDIH5}.}}
}

\begin{document}

\title{Automatic Differentiation for the Direct Minimization Approach to the Hartree--Fock Method}

\author[1]{Naruki Yoshikawa}
\author[2,3]{Masato Sumita}
\affil[1]{Department of Computer Science, University of Toronto}
\affil[2]{Center for Advanced Intelligence Project, RIKEN}
\affil[3]{International Center for Materials Nanoarchitectonics, National Institute for Materials Science}

\date{}

\maketitle

\begin{abstract}
Automatic differentiation has become an important tool for optimization problems in computational science, and it has been applied to the Hartree--Fock method. 
Although the reverse-mode automatic differentiation is more efficient than the forward-mode, eigenvalue calculation in the self-consistent field method has impeded the use of the reverse-mode automatic differentiation. Here, we propose a method to directly minimize Hartree--Fock energy under the orthonormality constraint of the molecular orbitals using reverse-mode automatic differentiation by avoiding eigenvalue calculation. According to our validation, the proposed method was more stable than the conventional self-consistent field method and achieved comparable accuracy.
\end{abstract}

\section{Introduction}
\thispagestyle{FirstPage}
Automatic differentiation (AD) is a technique to algorithmically calculate derivatives of a function without giving explicit forms of the derivatives. It has been widely accepted in machine learning; the backward propagation algorithm is a special case of reverse-mode automatic differentiation. Thanks to the recent advancements in software techniques, AD has been adopted in various fields of chemistry, such as molecular dynamics~\cite{schoenholz2020jax} and density functional theory (DFT)~\cite{kasim2021learning}, and several AD-based quantum chemistry software packages have been developed \cite{abbott2021arbitrary, kasim2021dqc}.
There are two types of AD: forward-mode and reverse-mode.
Reverse-mode AD is more efficient than forward-mode AD when the number of input variables is larger than that of output variables.

The Hartree--Fock method is an important approximation method in the electronic structure theory.
Conventionally, it results in an eigenvalue equation called the Roothaan equation by representing molecular orbitals as the linear combinations of atomic orbitals (LCAO).
The Roothaan equation is solved through the self-consistent field (SCF).
SCF requires calculating eigenvalues during its iterations, but AD of eigenvalue calculation has technical difficulty: reverse-mode AD cannot be applied in degenerated systems.
Previous work on the application of AD to the Hartree--Fock method~\cite{tamayo2018automatic} used the forward-mode AD for this reason.
However, reverse-mode AD is preferable because the Hartree–Fock method calculates one total energy value from multiple coefficients.

Another approach to calculating molecular orbital coefficients is the direct minimization of Hartree--Fock energy.
Based on the variational principle, the coefficients are optimized to minimize the total energy of a molecule.
The tricky part of this approach is the orthonormality condition of molecular orbitals.
An algorithm using QR decomposition to satisfy the orthonormality condition has been proposed~\cite{head1988optimization}, and it is implemented in a differentiable quantum chemistry library DQC~\cite{kasim2021dqc}.
The direct minimization approach is considered to be more robust~\cite{weber2008direct}, and it can be applied to large systems~\cite{OrderN}.

In this study, we investigated different approaches for the direct minimization of Hartree--Fock energy with the reverse-mode AD.
Combining AD, we implemented a curvilinear search algorithm using the Cayley transformation proposed by Wen and Yin~\cite{wen2013feasible}, which has been applied to DFT and outperformed SCF~\cite{zhang2014gradient}.
We also implemented the augmented Lagrangian method~\cite{powell1969method,hestenes1969multiplier} as a baseline for direct minimization.
Our approaches directly minimize the Hartree--Fock energy under the orthonormality constraint using a gradient obtained by reverse-mode AD without calculating eigenvalues.
We compared our AD method with the conventional SCF method accelerated by DIIS~\cite{pulay1980convergence, pulay1982improved}.
We showed that the curvilinear search method with AD is more stable than the traditional SCF while maintaining the same accuracy and confirmed that AD outperformed numerical differentiation in terms of calculation time.

\section{Method}
\subsection{Hartree--Fock method and constrained optimization}
The Hartree--Fock method is an approximation approach to solve the Schr\"{o}dinger equation for a molecular system.
In practice, the LCAO approximation is adopted.
In this approximation, the $i$-th molecular orbital $\psi_{i}$ is represented by
\begin{equation}
    \psi_{i} = \sum_p C_{pi} \phi_p,
\end{equation}
where $\phi_p$ is the $p$-th atomic orbital, and $C_{pi}$ is its coefficient.
Summation is taken over atomic orbitals.

The Roothaan equation~\cite{roothaan1951new} is a matrix formulation of the Hartree--Fock method within the LCAO approximation.
It is given by
\begin{equation}
    \mathbf{F}\mathbf{C} = \mathbf{S}\mathbf{C}\boldsymbol{\epsilon},
\end{equation}
where $\mathbf{F}$ is the Fock matrix, $\mathbf{C}$ is a matrix of LCAO coefficients, and $\mathbf{S}$ is the overlap matrix
\begin{equation}
    \mathbf{S}_{ij} = \int \phi_i^*(\mathbf{r})\phi_j(\mathbf{r}) d\mathbf{r}.
\end{equation}

The Roothaan equation is derived by applying the method of Lagrange multipliers to the minimization of the Hartree--Fock energy under the orthonomality condition
\begin{equation}
    \mathbf{C}^{\dagger}\mathbf{S}\mathbf{C} = \mathbf{I},\label{eq:orthonormality}
\end{equation}
where $\mathbf{I}$ is the identity matrix.

The Roothaan equation is usually solved iteratively since the Fock matrix $\mathbf{F}$ is a function of $\mathbf{C}$.
This iterative approach is called the self-consistent field (SCF).
However, the computational cost of the diagonalization in the iteration of SCF grows in the order of $n^3$, where $n$ is the size of coefficient matrix that depends on the number of electrons and the basis function.
In addition, the convergence of SCF is not theoretically guaranteed; they may oscillate between non-ground states even in simple molecules~\cite{cances2000convergence}.
Instead of solving the Roothaan equation iteratively, we can directly minimize Hartree--Fock energy under the orthonormality condition.

\subsection{Hartree--Fock energy}
We minimize the Hartree--Fock energy under the orthonormality constraint by adjusting the coefficients $\mathbf{C}$.
The restricted Hartree--Fock electronic energy $E$ as the function of the LCAO coefficients is given by~\cite{Szabo}
\begin{equation}
E(\mathbf{C}) = \frac{1}{2} \sum_{\mu}\sum_{\nu} {P}_{\nu\mu}({H}_{\mu\nu}^{\mathrm{core}} + {F}_{\mu\nu})\label{eq:energy},
\end{equation}
where $ {H}_{\mu\nu}^{\mathrm{core}} $ is a element of the core-Hamiltonian matrix $\mathbf{H}^{\mathrm{core}}$, ${F}_{\mu\nu}$ is a element of the Fock matrix $\mathbf{F}$, and ${P}_{\nu\mu}$ is a element of the charge-density bond-order matrix $\mathbf{P}$. 
Summation is taken over atomic orbitals.
Assuming that $N$ is even number of electrons, ${P}_{\nu\mu}$ is represented as
\begin{equation}
{P}_{\mu\nu} = 2 \sum_{a}^{N/2} \mathbf{C}_{\mu a}\mathbf{C}_{\nu a}.
\end{equation}
    
\subsection{Optimization with orthonormality constraints}
In this section, we consider an optimization problem with an orthogonality constraint

\begin{equation}
\begin{aligned}
&\text{minimize} && f(X)\\
&\text{subject to} && X^TMX=K,
\end{aligned}
\end{equation}
where $X \in \mathbb{R}^{n \times p}$, $M \in \mathbb{R}^{n \times n}$ is a symmetric positive definite matrix, and $K \in \mathbb{R}^{p \times p}$ is a nonsingular Hermitian matrix.
In the direct SCF problem, $f(X)$ is the energy function $E(\mathbf{C})$ defined by (\ref{eq:energy}), $X$ is the coefficients $\mathbf{C}$, $M$ is the overlap matrix $\mathbf{S}$, and $K$ is the identity matrix.
The overlap matrix $\mathbf{S}$ is a symmetric positive definite matrix and the identity matrix is a nonsingular Hermitian matrix, so Hartree--Fock energy minimization under the orthonormality constraint falls into this framework.

\subsubsection{Curvilinear search using Cayley transformation}
Here, we briefly explain the curvilinear search approach based on Cayley transformation~\cite{wen2013feasible}.
This approach minimizes the objective function along a descent path under the constraint.
Suppose a matrix $X$ satisfies $X^TMX = K$. We define
\begin{equation}
\begin{aligned}
    G &:= \left(\frac{\partial f(X)}{\partial X_{i, j}}\right),\\
    A &:= GX^TM - MXG^T.
\end{aligned}
\end{equation}
We will further define $Y(\tau)$ as the Cayley transformation
\begin{equation}
    Y(\tau) := \left(I + \frac{\tau}{2}AM\right)^{-1}\left(I - \frac{\tau}{2}AM\right)X.
\end{equation}
This transformation has several useful properties: $Y(\tau)^TMY(\tau) = X^TMX$, $Y(\tau)$ is smooth in $\tau$, and $\{Y(\tau)\}_{\tau \geq 0}$ is a descent path.
Therefore, we can run curvilinear search by choosing a proper step size $\tau$.
We used Barzilai-Borweing (BB) step size~\cite{barzilai1988two} for efficiency.
We define $X_k$ as the search point at iteration $k$. Then, the BB step size at iteration $k+1$ is
\begin{equation}
\begin{aligned}
    \tau_{k+1, 1} &:= \frac{\mathrm{tr}\left((S_k)^TS_k\right)}{\lvert\mathrm{tr}\left((S_k)^TY_k\right)\rvert}\quad  \text{or}\\[0.5em]
    \tau_{k+1, 2} &:= \frac{\lvert\mathrm{tr}\left((S_k)^TY_k\right)\rvert}{\mathrm{tr}\left((Y_k)^TY_k\right)},
\end{aligned}
\end{equation}
where $S_{k} = X_{k+1} - X_{k}$, $Y_{k} = G_{k+1} - G_k$.
Note that the definition of $Y_k$ is different from the original literature's definition ($Y_k = \nabla f(X_{k+1})-\nabla f(X_{k})$, where $\nabla f(X) = G - MXG^TXK^{-1}$), but our definition improved the performance of the numerical experiments.
The curvilinear search algorithm with BB steps is shown in Algorithm~\ref{alg:curvilinear}.
\begin{algorithm}[H]
\caption{Curvilinear search algorithm with BB steps}
\label{alg:curvilinear}
\begin{algorithmic}[1]
\renewcommand{\algorithmicrequire}{\textbf{Input:}}
\renewcommand{\algorithmicensure}{\textbf{Output:}}
\REQUIRE $f(X)$, $M$, $K$, $\tau, \tau_m, \tau_M > 0$, $\rho, \delta, \eta, \epsilon \in (0, 1)$, $k=0$
\ENSURE  $X$
\STATE initialize $X_0$ as a feasible point s.t. $X^TMX=K$
\STATE set $C_0 = f(X_0)$, $Q_0 = 1$
\STATE calculate $G_0$ and $A_0$
\WHILE{$\|A_k X_k\|>\epsilon$}
\WHILE{$f(Y_k(\tau)) \geq C_k - \rho \tau \|A_k\|_F^2$}
\STATE $\tau \leftarrow \delta \tau$
\ENDWHILE
\STATE $X_{k+1} \leftarrow Y_k(\tau)$, $Q_{k+1} \leftarrow \eta Q_k  + 1$ and \\$C_{k+1} \leftarrow (\eta Q_k C_k + f(X_{k+1}))/Q_{k+1}$
\STATE calculate $G_{k+1}$ and $A_{k+1}$
\STATE $\tau \leftarrow \mathrm{max}\left(\mathrm{min}(\tau_{k+1}, \tau_M), \tau_m\right)$, $k \leftarrow k+1$
\ENDWHILE
\end{algorithmic} 
\end{algorithm}
In step 10, we set $\tau_{k+1} = \tau_{k+1,1}$ if $k$ is even and $\tau_{k+1} = \tau_{k+1, 2}$ if $k$ is odd.
The convergence of this algorithm is guaranteed under reasonable conditions \cite{wen2013feasible}.

\subsubsection{Augmented Lagrangian method}
The augmented Lagrangian method~\cite{hestenes1969multiplier,powell1969method} is an algorithm to solve general constrained optimization problems. It has a wide range of applications, including quantum chemistry~\cite{sumita2021augmented}.
We use this algorithm as a baseline.
We consider the following optimization problem with an equality constraint
\begin{equation}
\begin{aligned}
    &\text{minimize} && f(X)\\
    &\text{subject to} && c(X) = 0,
\end{aligned}
\end{equation}

where $c(\mathbf{x}) = \|X^TMX-K\|_2$.
The augmented Lagrangian method iteratively solves this problem by reducing to an unconstrained optimization problem.
The algorithm is shown in Algorithm~\ref{alg:augmented}.

\begin{algorithm}[H]
\caption{Augmented Lagrangian method}
\label{alg:augmented}
\begin{algorithmic}[1]
\renewcommand{\algorithmicrequire}{\textbf{Input:}}
\renewcommand{\algorithmicensure}{\textbf{Output:}}
\REQUIRE objective function $f(X)$, constraint $c(X)$, tolerance $\varepsilon$
\ENSURE  $X$
\STATE initialize $\mu = 1$, $\lambda = 0$
\WHILE{$\lvert c(X) \rvert > \varepsilon$}
\STATE $X  \leftarrow \text{argmin} \: f(X) + \mu c(X)^2 + \lambda c(X)$
\STATE $\lambda \leftarrow \lambda + 2 \mu c(X)$ 
\STATE $\mu \leftarrow 2 \mu$
\ENDWHILE
\end{algorithmic} 
\end{algorithm}

We use Broyden–Fletcher–Goldfarb–Shanno (BFGS) \cite{nocedal2006numerical} to optimize the unconstrained optimization problem in the step 3.
The gradient obtained by automatic differentiation is used in BFGS.

\subsection{Automatic differentiation of Hartree--Fock energy}
Curvilinear search using Cayley transformation requires the gradient of the objective function to calculate the matrix G.
The augmented Lagrangian method also requires the Jacobian of the objective function for BFGS.
Since it is laborious and inefficient to implement the gradient of the energy function, we used AD.

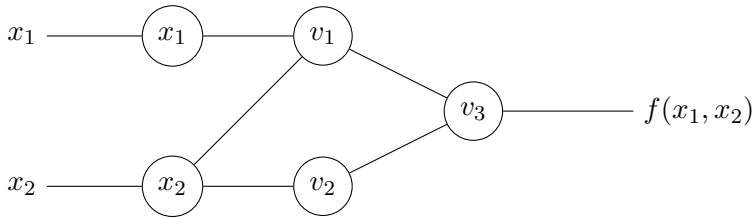
\begin{figure}[htbp]
\centering
\begin{tikzpicture}
    \node (X1) at (1, 3) {$x_1$};
    \node (X2) at (1, 1) {$x_2$};
    \node[circle, draw] (V0) at (3, 3) {$x_1$};
    \node[circle, draw] (V1) at (3, 1) {$x_2$};
    \node[circle, draw] (V2) at (5, 3) {$v_1$};
    \node[circle, draw] (V3) at (5, 1) {$v_2$};
    \node[circle, draw] (V4) at (7, 2) {$v_3$};
    \node (F) at (10, 2) {$f(x_1, x_2)$};
    \draw (X1) -- (V0);
    \draw (X2) -- (V1);
    \draw (V0) -- (V2);
    \draw (V1) -- (V2);
    \draw (V1) -- (V3);
    \draw (V3) -- (V4);
    \draw (V2) -- (V4);
    \draw (V4) -> (F);
\end{tikzpicture}
\caption{\label{fig:computation_graph} Computation graph for $f(x_1, x_2) = x_1x_2 + \ln (x_2)$.}
\end{figure}

AD is a technique to calculate the derivative of a function algorithmically based on the chain rule.
It is different from symbolic differentiation, which generates an explicit form of the derivative of a function.
It also differs from numerical differentiation, which estimates the value of derivatives from function values in different points.
AD algorithms are usually classified into forward-mode and reverse-mode.
The forward-mode AD calculates derivatives by applying the chain rule from input to output, while the reverse-mode AD calculates from output to input.
We illustrate the difference of these two ADs, using a simple example function, $f(x_1, x_2) = x_1x_2 + \ln (x_2)$.
The description here is based on a review paper~\cite{baydin2018automatic}.
A function can be visualized by a computation graph, a graph whose nodes correspond to variables and edges correspond to the dependencies of variables.
The computation graph for $f(x_1, x_2)$ is shown in Figure~\ref{fig:computation_graph}.

In the forward-mode AD, the intermediate variables $v_i$ and its derivative with respect to one target variable $x_j$ ($\dot{v_i} := \partial v_i / \partial x_j$) are calculated simultaneously. By applying the chain rule, the final derivative value can be computed. An example of forward-mode AD is shown in Table~\ref{tab:forward}.

\begin{table}
  \centering
  \renewcommand{\arraystretch}{1.2}
    \caption{An example of forward-mode AD. $y = f(x_1, x_2) = x_1x_2 + \ln (x_2)$ is evaluated at $(x_1, x_2)=(1, 2)$, and the derivative with respect to $x_1$ is calculated ($\dot{v} = \partial v/\partial x_1$).}
  \label{tab:forward}
  \begin{minipage}[t]{0.41\textwidth}
    \begin{tabularx}{\textwidth}[t]{p{0.5mm}p{0.8mm}p{18mm}@{}X}
      \toprule
      \multicolumn{4}{l}{Forward Primal Trace}\\
      \multirow{5}{1mm}{\begin{tikzpicture}\draw[->,>=triangle 60,thick](0,0)--(0,-3.8);\end{tikzpicture}} & $x_1$ & $=1$ &\\
      & $x_2$ & $=2$ &\\
      \cmidrule{2-4}
      & $v_1$ & $=x_1x_2$ & $=1 \times 2$\\
      & $v_2$ & $=\ln (x_2)$ & $=\ln 2$\\
      & $v_3$ & $=v_1 + v_2$ & $=2 + \ln 2$\\
      \cmidrule{2-4}
      & $y$ & $=v_3$ & $=2 + \ln 2$\\
      \bottomrule
    \end{tabularx}\vspace{1mm}
  \end{minipage}
  \begin{minipage}[t]{0.58\textwidth}
    \setlength{\fboxsep}{0pt}\colorbox{gray!20}{
       \begin{tabularx}{\textwidth}[t]{p{0.5mm}p{0.8mm}p{18mm}@{}X}
      \toprule
      \multicolumn{4}{l}{Forward Derivative Trace}\\
      \multirow{5}{1mm}{\begin{tikzpicture}\draw[->,>=triangle 60,thick](0,0)--(0,-3.8);\end{tikzpicture}} & $\dot{x}_1$ & $=1$ &\\
      & $\dot{x}_2$ & $=0$ &\\
      \cmidrule{2-4}
      & $\dot{v_1}$ & $=x_2$ & $=2$\\
      & $\dot{v_2}$ & $=0$ & \\
      & $\dot{v_3}$ & $=\dot{v_1} + \dot{v_2}$ & $=2$\\
      \cmidrule{2-4}
      & $\dot{y}$ & $=\dot{v_3}$ & $=2$\\
      \bottomrule
    \end{tabularx}}
  \end{minipage}
\end{table}

In the reverse-mode AD, the derivative is calculated in two phases: the forward calculation to calculate intermediate variables and record dependencies among them, and the reverse calculation to calculate derivative.
In the second phase, the adjoint of intermediate variable $\overline{v_i} = \partial y / \partial v_i$ is calculated from the output to input using the chain rule
\begin{align*}
    \frac{\partial y}{\partial x_i} = \sum_{j}\frac{\partial y}{\partial v_j}\frac{\partial v_j}{\partial x_i}
    = \sum_{j} \overline{v_j}\frac{\partial v_j}{\partial x_i}
\end{align*}
An example of reverse-mode AD is shown in Table~\ref{tab:reverse}.

\begin{table}
  \centering
  \renewcommand{\arraystretch}{1.2}
    \caption{An example of reverse-mode AD. $y = f(x_1, x_2) = x_1x_2 + \ln (x_2)$ is evaluated at $(x_1, x_2)=(1, 2)$ and the derivatives with respect to $x_1$, $x_2$ are calculated.}
  \label{tab:reverse}
  \begin{minipage}[t]{0.41\textwidth}
    \begin{tabularx}{\textwidth}[t]{p{0.5mm}p{0.8mm}p{18mm}@{}X}
      \toprule
      \multicolumn{4}{l}{Forward Primal Trace}\\
      \multirow{5}{1mm}{\begin{tikzpicture}\draw[->,>=triangle 60,thick](0,0)--(0,-3.8);\end{tikzpicture}} & $x_1$ & $=1$ &\\
      & $x_2$ & $=2$ &\\
      \cmidrule{2-4}
      & $v_1$ & $=x_1x_2$ & $=1 \times 2$\\
      & $v_2$ & $=\ln (x_2)$ & $=\ln 2$\\
      & $v_3$ & $=v_1 + v_2$ & $=2 + \ln 2$\\
      \cmidrule{2-4}
      & $y$ & $=v_3$ & $=2 + \ln 2$\\
      \bottomrule
    \end{tabularx}\vspace{1mm}
  \end{minipage}
  \begin{minipage}[t]{0.58\textwidth}
    \setlength{\fboxsep}{0pt}\colorbox{gray!20}{
       \begin{tabularx}{\textwidth}[t]{p{0.5mm}p{8mm}p{30mm}p{24mm}@{}X}
      \toprule
      \multicolumn{4}{l}{Reverse Derivative Trace}\\
      \multirow{5}{1mm}{\begin{tikzpicture}\draw[<-,>=triangle 60,thick](0,0)--(0,-3.8);\end{tikzpicture}} & $\overline{x}_1$ & $=\overline{v}_1 \frac{\partial v_1}{\partial x_1}$ & $= x_2$ & $=2$\\
      & $\overline{x}_2$ & $= \overline{v}_1 \frac{\partial v_1}{\partial x_2} + \overline{v}_2 \frac{\partial v_2}{\partial x_2}$ & $= x_1 + 1/x_2$ & $=3/2$\\
      \cmidrule{2-5}
      & $\overline{v}_1$ & $= \overline{v}_3 \frac{\partial v_3}{\partial v_1}$ & $= 1$\\
      & $\overline{v}_2$ & $= \overline{v}_3 \frac{\partial v_3}{\partial v_2}$ & $= 1$\\
      & & & \\
      \cmidrule{2-5}
    & $\overline{v_3}$ & $ = \overline{y}= \frac{\partial y}{\partial y}$ & $ = 1$\\
      \bottomrule
    \end{tabularx}}
  \end{minipage}
\end{table}
Forward-mode AD can calculate the derivative of all output variables for single input variable with constant factor additional time of the original function evaluation.
On the other hand, reverse-mode AD can calculate the derivative of single output variables for all input variables with constant factor additional time of the original function evaluation.
As a result, forward-mode is preferable when the number of output variables is larger than input variables; otherwise, the reverse-mode is preferable.
Further detail of automatic differentiation is described elsewhere~\cite{baydin2018automatic}.

In the Hartree--Fock energy calculation, the input variables are the LCAO coefficients of atomic orbitals and the output variable is the energy value. Thus, the number of input variables is larger than that of the output variable.
By constructing a computation graph for energy calculation, the derivative of energy can be obtained via AD.
However, previous work on AD for Hartree--Fock~\cite{tamayo2018automatic} used the forward-mode because their method depends on the derivatives of eigenvectors, and the reverse-mode AD of eigenvectors cannot be applied to systems with degenerated molecular orbitals.
Our direct minimization method does not use the derivatives of eigenvectors, so the reverse-mode AD is applicable.
We implemented energy calculation on JAX~\cite{jax2018github}, and we obtained the gradient of the energy function automatically by the \verb|grad()| function in JAX, which calculates the gradient by the reverse-mode AD.

\section{Results and Discussion}
We implemented direct minimization of the Hartree--Fock energy using the curvilinear search with Cayley transformation and the augmented Lagrangian method.
The core-Hamiltonian matrix and the Fock matrix are calculated by PySCF~\cite{PySCF}.
Gradients of the energy function is calculated by JAX~\cite{jax2018github}, which was set to 64-bit mode to obtain enough accuracy to preserve the constraint.
The parameter for the curvilinear search was as follows: $\tau = 1$, $\tau_m = 10^{-10}$, $\tau_{M} = 10^{10}$, $\rho=10^{-4}$, $\delta=0.1$, $\eta=0.5$, and $\epsilon=10^{-3}\:\text{or}\:10^{-6}$ depending on the molecular size.
$X_0$ is set to $S^{-1/2}$, the inverse square root of the overlap matrix, calculated by SciPy~\cite{scipy2020}.
As for the augmented Lagrangian method, the BFGS implemented by SciPy with $\varepsilon=10^{-6}$ was used in the internal minimization.
All experiments are conducted on a laptop with AMD Ryzen 7 PRO 4750U.
Our implementation is available at \url{https://github.com/n-yoshikawa/automatic-differentiation-SCF}.

\subsection{Diatomic molecules}
\begin{figure}[htbp]
\centering
\includegraphics[width=15cm]{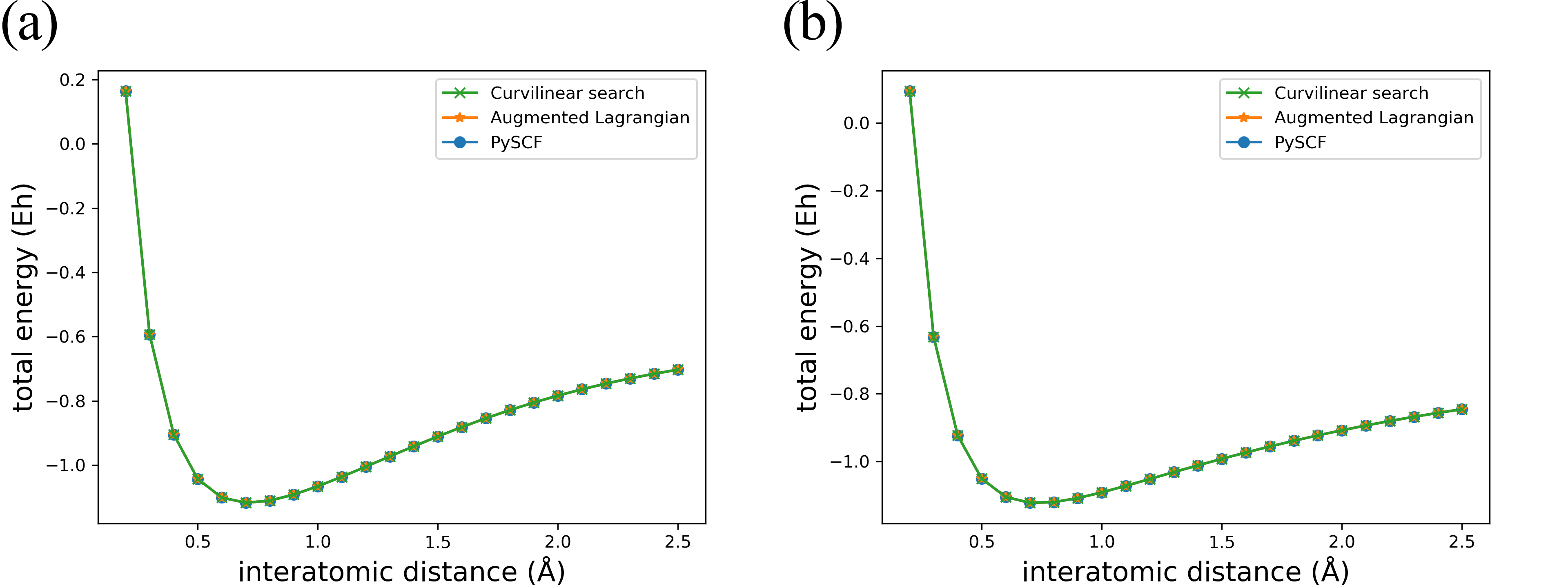}
\caption{\label{fig:h2} Energy curves of the hydrogen molecule (H$_2$) with the STO-3G basis set (a) and the 3-21G basis set (b). The the energy curves depicted by the curvilinear search method with AD show the same as the other two methods.  }
\end{figure}

\begin{figure}[htbp]
\centering
\includegraphics[width=15cm]{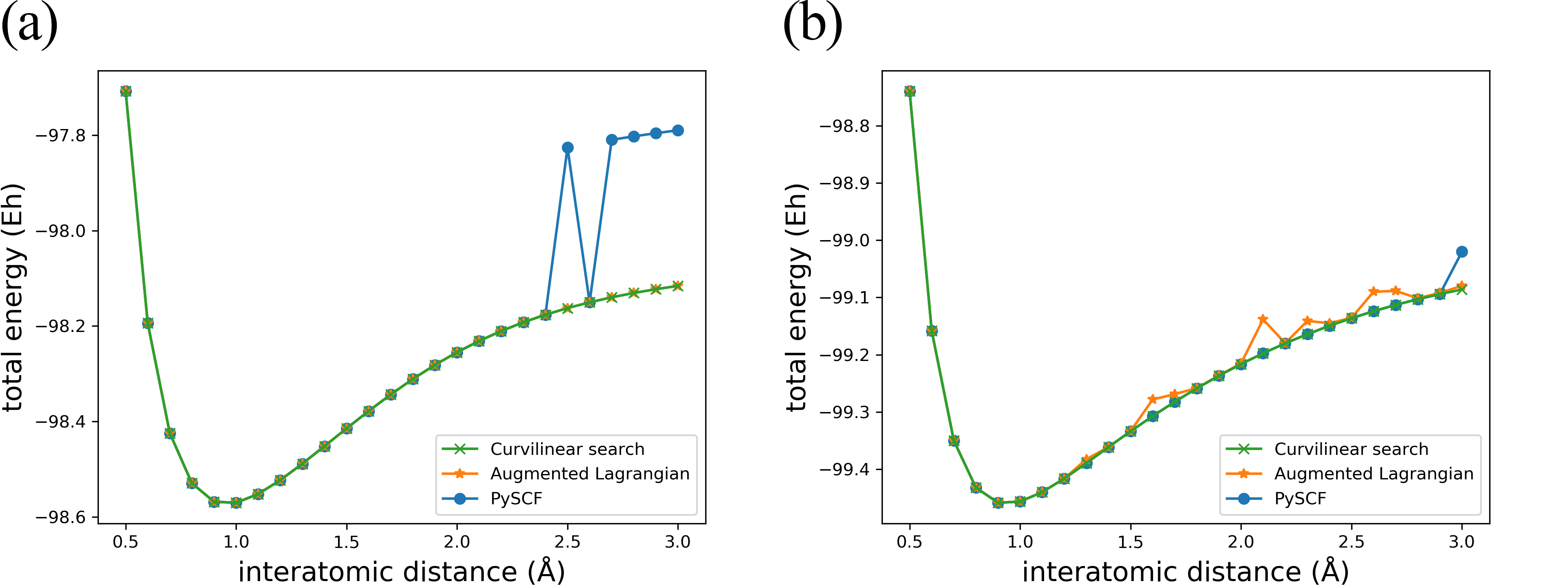}
\caption{\label{fig:hf} Energy curves of hydrogen fluoride (HF) with the STO-3G basis set (a) and the 3-21G basis set (b). The energy curves drawn by the curvilinear search method using AD were smooth and stable in all regions.}
\end{figure}

We calculated the total energies of diatomic atoms as the function of interatomic distance using the proposed methods.
For comparison, we calculated these energies using the conventional restricted Hartree--Fock (RHF) method implemented in PySCF.
PySCF uses DIIS~\cite{pulay1980convergence, pulay1982improved} to accelerate the convergence of SCF by default.
As indicated in Figure~\ref{fig:h2}, the calculated energies for the hydrogen molecule (H$_2$) were almost identical in the three methods.
The energy curve of the hydrogen fluoride (HF) molecule is shown in Figure~\ref{fig:hf}.
The curvilinear search method shows a stable potential energy curve even in the area of large interatomic distances, whereas the augmented Lagrangian method and RHF with PySCF did not converge at large interatomic distances.
These results indicate the effectiveness of the optimization with AD and curvilinear search.

\subsection{Polyatomic molecules}
\begin{figure}[htbp]
\centering
\includegraphics[width=15cm]{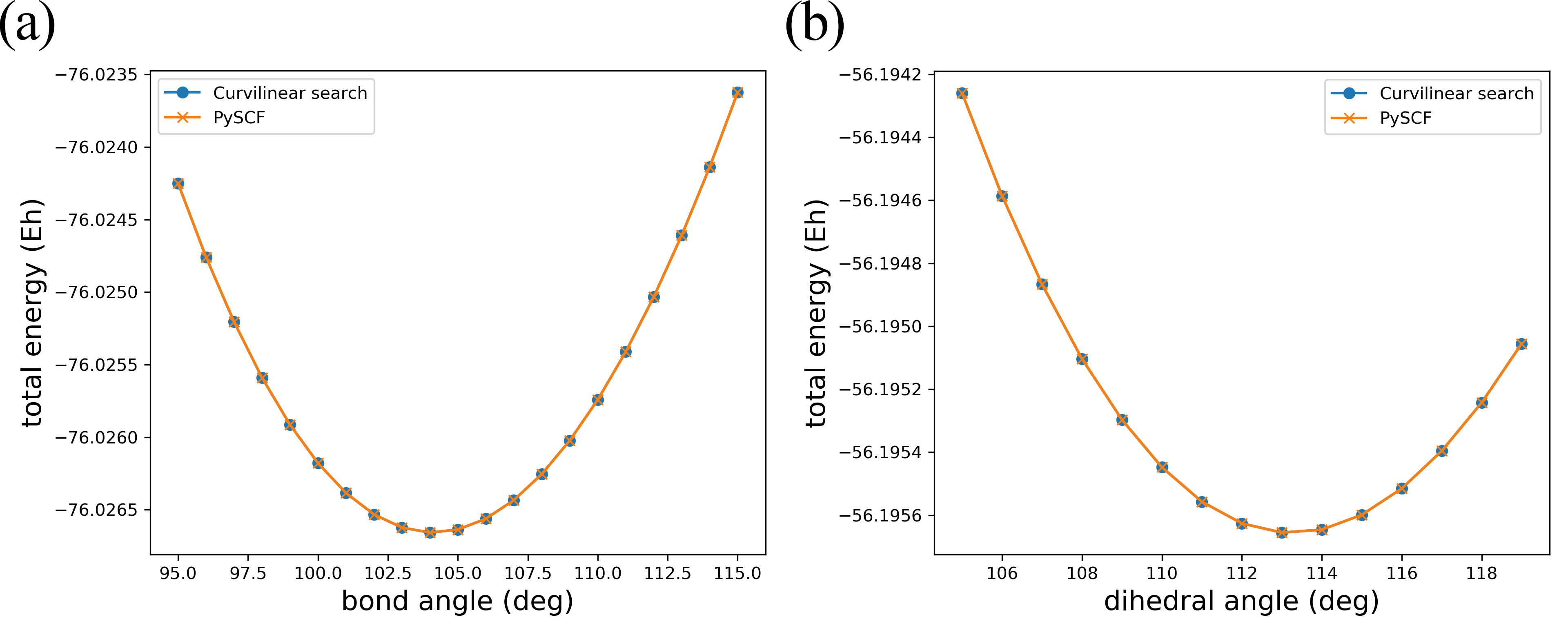}
\caption{\label{fig:angles} Energy curves of polyatomic molecules with the cc-pVDZ basis set. (a) Energy curve of H$_2$O in various bond angles. Bond length is fixed to 0.96 \AA. (b) Energy curve of NH$_3$ in various dihedral angles. Bond length is fixed to 1.01 \AA, and bond angle is fixed to 107$^{\circ}$.}
\end{figure}

We also applied the methods for calculating the total energies of some small polyatomic molecules. The energies of polyatomic molecules and their computational times are summarized in Table~\ref{tab:energy} and potential energy curves computed as the function of the bond angle of H$_2$O and the dihedral angle of NH$_3$ are shown in Figure~\ref{fig:angles}.
As shown in  Table~\ref{tab:energy}, SCF was the fastest method in all molecules, and the curvilinear search with Cayley transformation was second.
SCF usually converged in less than ten iterations in our results, while the other methods required hundreds of iterations to converge.
Because the Cayley transformation requires inverse matrix calculation, which requires $O(N^3)$ operations in many implementations, the larger number of iterations resulted in the slower calculation.
BFGS subroutine inside the augmented Lagrangian method also requires heavy calculations.
All three methods resulted in identical energy for all three methods in STO-3G basis set, but the augmented Lagrangian method resulted in higher energy in cc-pVDZ basis set.

\begin{table}[htbp]
\centering
\caption{\label{tab:energy}Calculated energies of molecules and times for computing them using the conventional SCF (SCF), the Cayley transformation-based method with AD (Cayley), and the augmented Lagrangian method with AD (AugLag).}
\scalebox{0.9}{
\begin{tabular}{llll}
\textrm{Molecule (Basis set)}&
\textrm{Method}&
\textrm{Energy ($E_{\rm{h}}$)}&
\textrm{Time (ms)}\\\hline
H$_2$O (STO-3G) & SCF & -74.957305 & 36.2\\
                & Cayley & -74.957305 & 687.5\\
                & AugLag & -74.957305 & 3817.1\\\hline
H$_2$O (cc-pVDZ) & SCF & -76.023527 & 47.6\\
               & Cayley & -76.023527 & 1772.3\\
               & AugLag & -75.908891 & 52368.5\\\hline
NH$_3$ (STO-3G) & SCF & -55.451235 & 38.9\\
               & Cayley & -55.451235 & 726.9\\
               & AugLag & -55.451235 & 3788.1\\\hline
NH$_3$ (cc-pVDZ) & SCF & -56.194061 & 70.8\\
                 & Cayley & -56.194061 & 3624.1\\
                 & AugLag & -56.121872 & 89459.8\\\hline
CH$_4$ (STO-3G) & SCF & -39.726699 & 36.2\\
                & Cayley & -39.726699 & 725.6\\
                & AugLag & -39.726699 & 4494.8\\\hline
CH$_4$ (cc-pVDZ) & SCF & -40.198710 & 76.8\\
                 & Cayley & -40.198710 & 5871.2\\
                 & AugLag & -36.507461 & 200083.9\\\hline
CHCH (STO-3G) & SCF & -75.855690 & 34.6\\
                & Cayley & -75.855690 & 853.8\\
                & AugLag & -75.855690 & 4878.0\\\hline     
CHCH (cc-pVDZ) & SCF & -76.824982 & 52.2\\
                & Cayley & -76.824982 & 84550.6\\
                & AugLag & -75.348432 & 270444.5\\\hline
CH$_2$CH$_2$ (STO-3G) & SCF & -77.072653 & 35.6\\
                & Cayley & -77.072653 & 788.7\\
                & AugLag & -77.072653 & 4649.6\\\hline
CH$_2$CH$_2$ (cc-pVDZ) & SCF & -78.039252 & 250.2\\
                & Cayley & -78.039252 & 46216.7\\
                & AugLag & -67.093328 & 1124327.1\\\hline
CH$_3$CH$_3$ (STO-3G) & SCF & -76.566573 & 236.2\\
                & Cayley & -76.566573 & 14784.8\\
                & AugLag & -76.173236 & 4494.8\\\hline
CH$_3$CH$_3$ (cc-pVDZ) & SCF & -77.680403 & 281.8\\
                & Cayley & -77.680403 & 327998.1\\
                & AugLag & -59.996150 & 3030482.4\\\hline
CH$_3$F (cc-pVDZ) & SCF & -139.044219 & 253.6\\
                & Cayley & -139.044219 & 9140.6\\
                & AugLag & -131.036979 & 810791.3\\\hline
CH$_2$O (cc-pVDZ) & SCF & -113.875243 & 66.9\\
                & Cayley & -113.875243 & 8575.4\\
                & AugLag & -110.082746 & 408750.9\\\hline
\end{tabular}
}
\end{table}

\subsection{Effect of automatic differentiation}
AD provides a numerically precise gradient of the electronic structure theory like the Hartree--Fock energy, whose analytical gradient is sometimes tedious to derive.
Numerical differentiation based on finite difference is another approach to numerically calculate the gradient of functions.
It is easy to implement but highly susceptible to rounding errors, and it requires many function calls, at least as many as the number of variables~\cite{iri1988automatic}.
We compared AD and numerical differentiation by replacing the automatically derived gradient function in Cayley transformation with the finite difference approximation of the gradient.
Forward finite difference $(f(\mathbf{x} + \boldsymbol\varepsilon_i) - f(\mathbf{x})) / \varepsilon$ is used as an approximation of the $i$-th element of the gradient.
Here, $\boldsymbol\varepsilon_i$ is a vector whose $i$-th element is set to a small value $\varepsilon$ and other elements are zero.
We used SciPy's \verb|approx_fprime()| function for numerical differentiation, and set $\varepsilon = 1.49\times10^{-8}$, which is the SciPy's default.

\begin{figure}[htbp]
\centering
\includegraphics[width=15cm]{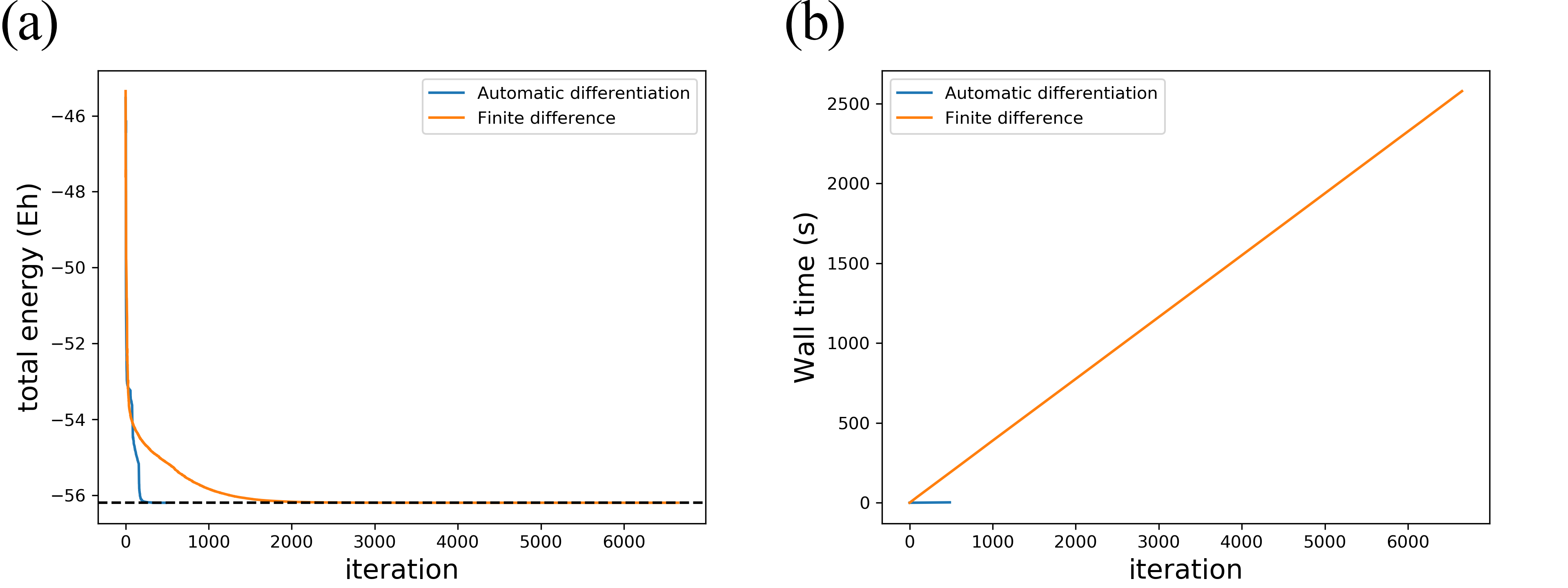}
\caption{\label{fig:nh3_itr} Energy calculation of NH$_3$ with the cc-pVDZ basis set. (a) Energy change in each iteration. Black dashed line indicates the energy calculated by SCF. AD (blue line) converges much faster than FD (orange line). (b) Wall time in each iteration. FD (orange line) required more time to converge than AD (blue line) because of the longer execution time per iteration and larger number of iterations required for convergence.}
\end{figure}

Table~\ref{tab:advsnd} shows the comparison of AD and numerical differentiation by finite difference (FD) in small polyatomic molecules.
Both AD and FD resulted in the same energy, but AD converged faster than FD.
To analyze the difference in speed, we examined the convergence of the algorithm.
Figure~\ref{fig:nh3_itr} shows the energy change and wall time at each iteration in the energy calculation for the NH$_3$ molecule.
FD required more iterations to converge because the energy decrease per iteration was smaller than AD possibly due to the inaccurate gradient estimate.
In addition, the wall time per iteration was larger in FD than AD because FD requires more function evaluations to calculate a gradient.

\begin{figure}[htbp]
\centering
\includegraphics[width=15cm]{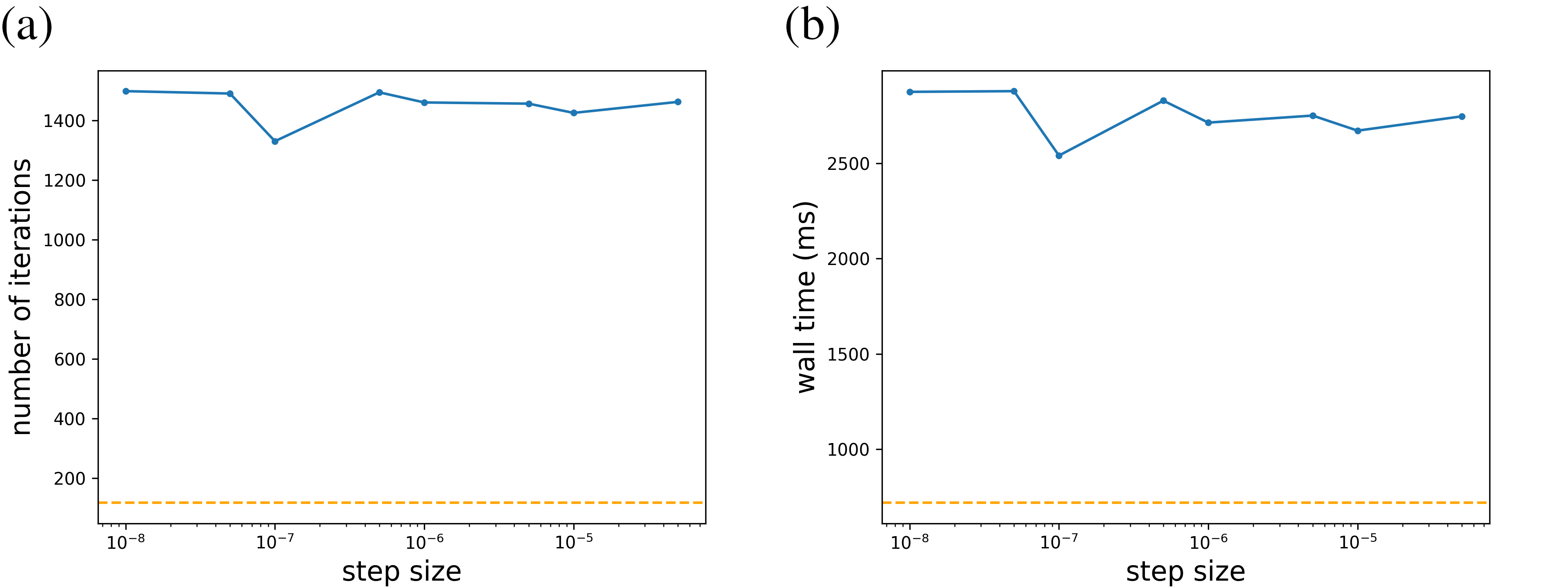}
\caption{\label{fig:stepsize} Effect of step size for energy calculation of H$_2$O with the STO-3G basis set using finite difference. Note that the calculation did not converge when step size is larger than $5 \times 10^{-5}$. (a) The number of iterations required for convergence in each step size. The performance was different, depending on step size, but no value outperformed automatic differentiation (orange). (b) Wall time required for convergence in each step. The same trend as iteration number was obtained.}
\end{figure}

The accuracy of gradient estimation evaluated by finite difference depends on the step size.
Figure~\ref{fig:stepsize} shows the dependence of step size for energy calculation of H$_2$O.
The step size affects the performance of algorithm in terms of required number of iterations for convergence and wall time, but AD outperforms FD in all settings.
Moreover, AD does not suffer from choosing optimal step size.
Therefore, we can conclude that the calculating gradient by AD contributes to the overall performance.

\begin{table}[htbp]
\centering
\caption{\label{tab:advsnd}Comparison of automatic differentiation (AD) and numerical differentiation with finite difference (FD). The calculated energy, total execution time, and the number of iterations until reaching convergence are shown. AD outperforms FD in terms of shorter execution time and less number of iterations in all settings.}
\begin{tabular}{lllll}
Molecule (Basis set) & Method & Energy ($E_{\rm{h}}$)&Time (ms)&Iterations\\\hline
H$_2$O (STO-3G) & AD & -74.957305 & 719.1 & 117\\
               & FD & -74.957305 & 2767.4 & 1414\\\hline
H$_2$O (3-21G) & AD & -75.584803 & 1676.2 & 849\\
               & FD & -75.584803 & 18390.4 & 2602\\\hline
H$_2$O (cc-pVDZ) & AD & -76.023527 & 1772.3 & 342\\
                 & FD & -76.023527 & 536394.0 & 3445\\\hline
NH$_3$ (STO-3G) & AD & -55.451235 & 726.9 & 124\\
                & FD & -55.451235 & 1600.5 & 431\\\hline
NH$_3$ (3-21G) & AD & -55.872058 & 869.1 & 219\\
               & FD & -55.872058 & 16571.5 & 1965\\\hline
NH$_3$ (cc-pVDZ) & AD & -56.194061 & 3624.1 & 413\\
                 & FD & -56.194061 & 3235168.0 & 7325\\\hline
CH$_4$ (STO-3G) & AD & -39.726699 & 725.6 & 110\\
                & FD & -39.726699 & 2456.9 & 666\\\hline
CH$_4$ (3-21G) & AD & -39.976739 & 1466.1 & 537\\
               & FD & -39.976739 & 38680.2 & 3153\\\hline
CH$_4$ (cc-pVDZ) & AD & -40.198710 & 5871.2 & 471\\
                 & FD & -40.198710 & 4681733.6 & 4259\\\hline
\end{tabular}
\end{table}


\section{Conclusions}
We proposed a constrained optimization method using AD as one of the direct minimization methods for the Hartree--Fock method.
We applied the method to compute the total energies of small molecules and describe the potential energy curves as the function of the internal degrees of freedom of some molecules.
Our method gave almost identical energies for small molecules as those obtained by the SCF with iterative diagonalization. 
Furthermore, potential energy curves along interatomic distances, bond angle, and dihedral angle were stable, and the calculation converged in some cases where SCF with DIIS did not converge. 
Therefore, we conclude that the proposed method can be an alternative to the traditional SCF methods.
Also, we verified that AD is advantageous over numerical differentiation in terms of speed.
However, the calculation speed of the proposed method was not as fast as the conventional SCF method, and applicability to larger molecules was limited.
Further investigation for acceleration technique, such as finding a good initial guess, is necessary to make our approach have a practical advantage.
A previous work~~\cite{dai2017conjugate} reported that conjugate gradient can outperform gradient method.
Combination of AD and conjugate gradient is another possible future work.

\subsection*{Acknowledgments}
We thank Rodrigo Vargas and Al\'{a}n Aspuru-Guzik for useful discussions.

\bibliographystyle{unsrt}
\bibliography{reference}

\end{document}